\documentstyle[12pt]{article}
\begin{document}
\begin{center}{\large{\bf Conformal Form of Pseudo-Riemannian Metrics by Normal Coordinate Transformations II }}
\end{center}
\vspace*{1.5cm}
\begin{center}
A. C. V. V. de Siqueira
$^{*}$ \\
Departamento de Educa\c{c}\~ao\\
Universidade Federal Rural de Pernambuco \\
52.171-900, Recife, PE, Brazil.\\
\end{center}
\vspace*{1.5cm}
\begin{center}{\bf Abstract}

In this paper, we have reintroduced a new approach to conformal
geometry developed and presented in two previous papers,
in which we show that all n-dimensional pseudo-Riemannian
metrics are conformal to a flat n-dimensional manifold as well as
an n-dimensional manifold of constant curvature when Riemannian
normal coordinates are well-behaved in the origin and in their
neighborhood. This was based on an approach developed by French
mathematician Elie Cartan. As a consequence of geometry, we have
reintroduced the classical and quantum angular momenta of a
particle and present new interpretations. We also show that
all n-dimensional pseudo-Riemannian metrics can be embedded in a
hyper-cone of a flat n+2-dimensional manifold.
  \end{center}
\vspace{3cm}
${}^*$ E-mail: acvvs@ded.ufrpe.br
\newpage
\section{Introduction}
$         $
In this paper, we reintroduce a new approach to conformal
geometry developed and presented in two previous papers,
\cite{1}, \cite{2}. This was based on an approach developed
by French mathematician Elie Cartan,\cite{3}, \cite{4}, \cite{5}.
Some classical and quantum results are reintroduced from a new viewpoint.
\newline
This paper is organized as follows. In Sec. $2$, we present normal
coordinates and elements of differential geometry. In Sec. $3$, we
show that, in normal coordinates, all n-dimensional pseudo-Riemannian
metrics that are well-behaved in origin and in their neighborhood
are conformal to a flat n-dimensional manifold and an n-dimensional
manifold of constant curvature. In Sec. $4$, this result is used in the
Cartan solution for a space of constant curvature. In Sec. $5$,
we present more differential geometry, introducing normal tensors
to build the Cartan solution for a general pseudo-Riemannian metric.
In Sec. $6$, we make an embedding of all n-dimensional pseudo-Riemannian
manifolds of constant curvature in a flat n+1-dimensional manifold,
obtaining the quantum angular momentum operator of a particle as a
consequence of geometry. In Sec. $7$, a new geometric postulate is
announced and some new classical physical principles are developed.
In Sec. $8$, we make an embedding of all n-dimensional pseudo-Riemannian
metrics that obey previously presented conditions into a hyper-cone
of a flat n+2-dimensional space.
\renewcommand{\theequation}{\thesection.\arabic{equation}}
\section{\bf  Normal Coordinates}
$         $
 \setcounter{equation}{0}
 $         $
 \setcounter{equation}{0}
 $         $
In this section, we briefly present normal coordinates and review
some elements of differential geometry for an n-dimensional
pseudo-Riemannian  manifold, \cite{3}, \cite{4}, \cite{5}.
\newline
Let us consider the line element
\begin{equation}
 ds^2= G_{\Lambda\Pi}du^{\Lambda}du^{\Pi},
\end{equation}
with
\begin{equation}
G_{\Lambda\Pi}=E_{\Lambda}^{(\mathbf{A})}E_{\Pi}^{(\mathbf{B})}\eta_{(\mathbf{A})(\mathbf{B})},
\end{equation}
in which $ \eta_{(\mathbf{A})(\mathbf{B})}$ and $
E_{\Lambda}^{(\mathbf{A})}$ are flat metric and vielbein
components, respectively.
\newline
We choose each
 $ \eta_{(\mathbf{A})(\mathbf{B})}$ as plus or minus Kronecker's delta function.
Let us give the 1-form  $\omega^{(\mathbf{A})} $ by
\begin{equation}
\omega^{(\mathbf{A})}= du^{\Lambda} E_{\Lambda}^{(\mathbf{A})}.
\end{equation}
We now define Riemannian normal coordinates by
\begin{equation}
  u^{\Lambda}=v^{\Lambda}t,
\end{equation}
then
\begin{equation}
 du^{\Lambda}=v^{\Lambda}dt+tdv^{\Lambda}.
\end{equation}
Placing (2.5) in (2.3)
\begin{equation}
  \omega^{(\mathbf{A})}= tdv^{\Lambda}
  E_{\Lambda}^{(\mathbf{A})}+dtv^{\Lambda}E_{\Lambda}^{(\mathbf{A})}.
\end{equation}
Let us define
\begin{equation}
 z^{(\mathbf{A})}=v^{\Lambda}E_{\Lambda}^{(\mathbf{A})},
\end{equation}
so that
\begin{equation}
\omega^{(\mathbf{A})}=dtz^{(\mathbf{A})}+tdz^{(\mathbf{A})}
+tE^{\Pi(\mathbf{A})}\frac{\partial{E_{\Pi(\mathbf{B})}}}{\partial{z^{(\mathbf{C})}}}z^{(\mathbf{B})}dz^{(\mathbf{C})}.\\
\end{equation}
We now make
\begin{equation}
 A^{(\mathbf{A})_{(\mathbf{B})(\mathbf{C})}}=tE^{\Pi(\mathbf{A})}\frac{\partial{E_{\Pi(\mathbf{B})}}}{\partial{z^{(\mathbf{C})}}},
\end{equation}
then
\begin{equation}
\varpi^{(\mathbf{A})}=
tdz^{(\mathbf{A})}+A^{({A})_{(\mathbf{B})(\mathbf{C})}}z^{(\mathbf{B})}dz^{(\mathbf{C})},
\end{equation}
with
\begin{equation}
\omega^{(\mathbf{A})}=dtz^{(\mathbf{A})}+\varpi^{(\mathbf{A})}.
\end{equation}
At $t=0$, we have
\begin{equation}
 A^{({A})_{(\mathbf{B})(\mathbf{C})}}(t=0,z^{(\mathbf{D})})=0,
\end{equation}
\begin{equation}
\varpi^{(\mathbf{A})}(t=0,z^{(\mathbf{D})})=0 ,
\end{equation}
and
\begin{equation}
\omega^{(\mathbf{A})}(t=0,z^{(\mathbf{D})})=dtz^{(\mathbf{A})} .
\end{equation}
We conclude that $\omega^{(\mathbf{A})}$ is the one-form associated
to the normal coordinate $u^{\Lambda}$, $z^{(\mathbf{A})}$ is
associated to the local coordinate $v^{\Lambda}$ of a local basis
and $\varpi^{(\mathbf{A})}$ is the one-form associated to the
one-form $dz^{(\mathbf{A})}$.
\newline
In an n+1-manifold, consider a
coordinate system given by $(t,z^{(\mathbf{A})})$. For each value of
t, we have a hyper-surface in which $dt=0$ on each. We are
interested in the hyper-surface with $t=1$. On this hyper-surface, we
verify the following equality:
\begin{equation}
\omega^{(\mathbf{A})}(t=1,z)=\varpi^{(\mathbf{A})}(t=1,z).
\end{equation}
Consider the following expression on a vielbein basis:
\begin{equation}
d\omega^{(\mathbf{A})}=-\omega^{(\mathbf{A})}_{(\mathbf{B})}\wedge\omega^{(\mathbf{B})}.
\end{equation}
The expression is invariant by coordinate transformations.
\newline
Now consider the map $\Phi$ between two manifolds M and N,
\newline
and consider two subsets, U of M and V of N. Then,
\begin{equation}
\Phi:U\longrightarrow V.
\end{equation}
We now define pull-back as follows: \cite{4},
\begin{equation}
\Phi^\ast:F^p(V)\longrightarrow F^p(U),
\end{equation}
so that $ \Phi^\ast $ sends p-forms to other p-forms.
\newline
It is well known that the exterior derivative commutes with
pull-back, so that
\begin{equation}
\Phi^\ast(d\omega^{(\mathbf{A})}_{(\mathbf{B})})=d\Phi^\ast(\omega^{(\mathbf{A})}_{(\mathbf{B})}),
\end{equation}
and
\begin{equation}
\Phi^\ast(d\omega^{(\mathbf{A})})=d\Phi^\ast(\omega^{(\mathbf{A})}).
\end{equation}
We also have
\begin{equation}
\Phi^\ast(\omega^{(\mathbf{A})}_{(\mathbf{B})}\wedge\omega^{(\mathbf{B})})=
\Phi^\ast(\omega^{(\mathbf{A})}_{(\mathbf{B})})\wedge\Phi^\ast(\omega^{(\mathbf{B})}).
\end{equation}
The equation (2.11) can be seen as a pull-back,
\begin{equation}
\Phi^\ast(\omega^{(\mathbf{A})})=dtz^{(\mathbf{A})}+\varpi^{(\mathbf{A})}.
\end{equation}
By a simple calculation, it can be shown that
\begin{equation}
\Phi^\ast(\omega^{(\mathbf{A})}_{(\mathbf{B})})=
\varpi^{(\mathbf{A})}_{(\mathbf{B})}.
\end{equation}
Note that $dt=0$ for $\varpi^{(\mathbf{A})}$ and
$\varpi^{(\mathbf{A})}_{(\mathbf{B})}$.
\newpage
By the exterior derivative of (2.22), we obtain
\begin{eqnarray}
\nonumber
d(\Phi^\ast(\omega^{(\mathbf{A})}))=d(dtz^{(\mathbf{A})}+\varpi^{(\mathbf{A})})=dz^{(\mathbf{A})}\wedge(dt)\\
\nonumber+dt\wedge\frac{\partial(\varpi^{(\mathbf{A})})}{\partial(t)}\\
\end{eqnarray}
\newline
+ terms not involving $dt$.
\vspace{1cm}
\newline
Making a pull-back of (2.16) and using (2.21), we have
\begin{equation}
\Phi^\ast(d\omega^{(\mathbf{A})})=\Phi^\ast(-\omega^{(\mathbf{A})}_{(\mathbf{B})}\wedge\omega^{(\mathbf{B})})=
-\Phi^\ast(\omega^{(\mathbf{A})}_{(\mathbf{B})})\wedge\Phi^\ast(\omega^{(\mathbf{B})}).
\end{equation}
Using (2.20), (2.23), (2.24) and (2.25), we obtain
\begin{equation}
\frac{\partial\varpi^{(\mathbf{A})}}{\partial{t}}=dz^{(\mathbf{A})}+
\varpi^{(\mathbf{A})}_{(\mathbf{B})}z^{(\mathbf{D})}.
\end{equation}
By a similar procedure to (2.19) and using the Cartan
second structure equation, we obtain the following result:
\begin{equation}
\frac{\partial\varpi_{(\mathbf{A})(\mathbf{B})}}{\partial{t}}=
R_{(\mathbf{A})(\mathbf{B})(\mathbf{C})(\mathbf{D})}z^{(\mathbf{C})}\varpi^{(\mathbf{A})}.
\end{equation}
Making a new partial derivative of (2.26), two partial derivatives
of (2.10), comparing the results and using (2.27), we have the
following equation:
\begin{equation}
\frac{\partial^2A_{(\mathbf{A}){(\mathbf{C})(\mathbf{D})}}}{\partial{t^2}}=
tz^{(\mathbf{B})}R_{(\mathbf{A})(\mathbf{B})(\mathbf{C})(\mathbf{D})}+
z^{(\mathbf{L})}z^{(\mathbf{M})}R_{(\mathbf{A})(\mathbf{L})(\mathbf{M})(\mathbf{N})}
A_{(\mathbf{P}){(\mathbf{C})(\mathbf{D})}}\eta^{(\mathbf{N})(\mathbf{P})}.
\end{equation}
Rewriting (2.28) with indices (C) and (D) permuted, we obtain
the result
\begin{equation}
\frac{\partial^2A_{(\mathbf{A}){(\mathbf{D})(\mathbf{C})}}}{\partial{t^2}}=
tz^{(\mathbf{B})}R_{(\mathbf{A})(\mathbf{B})(\mathbf{D})(\mathbf{C})}+
z^{(\mathbf{L})}z^{(\mathbf{M})}R_{(\mathbf{A})(\mathbf{L})(\mathbf{M})(\mathbf{N})}
A_{(\mathbf{P}){(\mathbf{D})(\mathbf{C})}}\eta^{(\mathbf{N})(\mathbf{P})}.
\end{equation}
Adding (2.28) and (2.29) and using the curvature symmetries, we have
\begin{equation}
A_{(\mathbf{A}){(\mathbf{C})(\mathbf{D})}}+A_{(\mathbf{A}){(\mathbf{D})(\mathbf{C})}}=0,
\end{equation}
which is true for all t.
\newline
Then,
\begin{equation}
A_{(\mathbf{A}){(\mathbf{C})(\mathbf{D})}}=-A_{(\mathbf{A}){(\mathbf{D})(\mathbf{C})}},
\end{equation}
such that we can rewrite (2.10) as
\begin{equation}
\varpi^{(\mathbf{A})}= tdz^{(\mathbf{A})}+
\frac{1}{2}A^{({A})_{(\mathbf{B})(\mathbf{C})}}(z^{(\mathbf{B})}dz^{(\mathbf{C})}-z^{(\mathbf{C})}dz^{(\mathbf{B})}).
\end{equation}
Let us define
\begin{equation}
A_{(\mathbf{A}){(\mathbf{C})(\mathbf{D})}}=z^{(\mathbf{B})}B_{(\mathbf{A}){(\mathbf{B})(\mathbf{C})(\mathbf{D})}}.
\end{equation}
The following result is obtained by placing (2.33) in (2.28):
\begin{equation}
\frac{\partial^2B_{(\mathbf{A}){(\mathbf{B})(\mathbf{C})(\mathbf{D})}}}{\partial{t^2}}=
tR_{(\mathbf{A})(\mathbf{B})(\mathbf{C})(\mathbf{D})}+
z^{(\mathbf{L})}z^{(\mathbf{M})}R_{(\mathbf{A})(\mathbf{B})(\mathbf{L})(\mathbf{N})}
B_{(\mathbf{P}){(\mathbf{M})(\mathbf{C})(\mathbf{D})}}\eta^{(\mathbf{N})(\mathbf{P})}.
\end{equation}
We now rewrite (2.34) as
\begin{equation}
\frac{\partial^2B_{(\mathbf{B}){(\mathbf{A})(\mathbf{C})(\mathbf{D})}}}{\partial{t^2}}=
tR_{(\mathbf{B})(\mathbf{A})(\mathbf{C})(\mathbf{D})}+
z^{(\mathbf{L})}z^{(\mathbf{M})}R_{(\mathbf{B})(\mathbf{A})(\mathbf{L})(\mathbf{N})}
B_{(\mathbf{P}){(\mathbf{M})(\mathbf{C})(\mathbf{D})}}\eta^{(\mathbf{N})(\mathbf{P})}.
\end{equation}
Adding (2.34) and (2.35) and using the curvature symmetries, we
obtain the solution
\begin{equation}
B_{(\mathbf{A}){(\mathbf{B})(\mathbf{C})(\mathbf{D})}}+B_{(\mathbf{B}){(\mathbf{A})(\mathbf{C})(\mathbf{D})}}=const.
\end{equation}
for all t.
\newline
We use (2.12) and (2.33) in (2.36) to obtain
\begin{equation}
B_{(\mathbf{A}){(\mathbf{B})(\mathbf{C})(\mathbf{D})}}+B_{(\mathbf{B}){(\mathbf{A})(\mathbf{C})(\mathbf{D})}}=0.
\end{equation}
For future use, we present the line element on
the hyper-surface
\begin{equation}
 ds'^2=\eta_{(\mathbf{A})(\mathbf{B})}\varpi^{(\mathbf{A})}\varpi^{(\mathbf{B})}.
\end{equation}
From (2.31), (2.33) and (2.37), we conclude that
$B_{(\mathbf{A}){(\mathbf{B})(\mathbf{C})(\mathbf{D})}} $ has the
same symmetries as the Riemann curvature tensor
\begin{equation}
B_{(\mathbf{A}){(\mathbf{B})(\mathbf{C})(\mathbf{D})}}=-B_{(\mathbf{B}){(\mathbf{A})(\mathbf{C})(\mathbf{D})}}=
-B_{(\mathbf{A}){(\mathbf{B})(\mathbf{D})(\mathbf{C})}}.
\end{equation}
 Using (2.31) and (2.33), we have
\begin{eqnarray}
\nonumber A_{(\mathbf{A}){(\mathbf{C})(\mathbf{D})}}dz^{(\mathbf{A})}z^{(\mathbf{C})}dz^{(\mathbf{D})}=\\
\nonumber +\frac{1}{4}B_{(\mathbf{A}){(\mathbf{B})(\mathbf{C})(\mathbf{D})}}.\\
\nonumber .(z^{(\mathbf{B})}dz^{(\mathbf{A})}-z^{(\mathbf{A})}dz^{(\mathbf{B})}).\\
\nonumber .(z^{(\mathbf{C})}dz^{(\mathbf{D})}-z^{(\mathbf{D})}dz^{(\mathbf{C})}).\\
\end{eqnarray}
We now construct the line element of the hyper-surface. With
direct use of (2.32) and (2.40)in (2.38), we have
\begin{eqnarray}
\nonumber ds'^2=t^2\eta_{(\mathbf{A})(\mathbf{B})}dz^{(\mathbf{A})}dz^{(\mathbf{B})}+\\
 \nonumber+\frac{1}{2}\{\frac{1}{2}t\epsilon_{(\mathbf{B})}B_{(\mathbf{A}){(\mathbf{B})(\mathbf{C})(\mathbf{D})}}+\\
 \nonumber+\eta^{(\mathbf{M})(\mathbf{N})}A_{(\mathbf{M}){(\mathbf{B})(\mathbf{A})}}A_{(\mathbf{N}){(\mathbf{C})(\mathbf{D})}}\}.\\
\nonumber.(z^{(\mathbf{B})}dz^{(\mathbf{A})}-z^{(\mathbf{A})}dz^{(\mathbf{B})})(z^{(\mathbf{C})}dz^{(\mathbf{D})}-z^{(\mathbf{D})}dz^{(\mathbf{C})}).\\
\end{eqnarray}
The line element of the manifold and the hyper-surface are equal
at $ t=1 $, where  $u^{\Lambda}=v^{\Lambda} $,
\begin{equation}
ds^2=ds'^2,
\end{equation}
and
\begin{eqnarray}
\nonumber ds^2=\eta_{(\mathbf{A})(\mathbf{B})}dz^{(\mathbf{A})}dz^{(\mathbf{B})}+\\
 \nonumber+\frac{1}{2}\{\frac{1}{2}\epsilon_{(\mathbf{B})}B_{(\mathbf{A}){(\mathbf{B})(\mathbf{C})(\mathbf{D})}}+\\
 \nonumber+\eta^{(\mathbf{M})(\mathbf{N})}A_{(\mathbf{M}){(\mathbf{B})(\mathbf{A})}}A_{(\mathbf{N}){(\mathbf{C})(\mathbf{D})}}\}.\\
\nonumber.(z^{(\mathbf{B})}dz^{(\mathbf{A})}-z^{(\mathbf{A})}dz^{(\mathbf{B})})(z^{(\mathbf{C})}dz^{(\mathbf{D})}-z^{(\mathbf{D})}dz^{(\mathbf{C})}).\\
\end{eqnarray}
Note that (2.43) is not an approximation of (2.1); the equations are equal.
\newpage
\renewcommand{\theequation}{\thesection.\arabic{equation}}
\section{\bf Conformal Form of Riemannian Metrics }
 \setcounter{equation}{0}
$         $
 \setcounter{equation}{0}
 $         $
Sometimes, it is possible to write the metric (2.43) in the particular form (3.1), as follows:
\begin{eqnarray}
\nonumber ds^2=\eta_{(\mathbf{a})(\mathbf{b})}dz^{(\mathbf{a})}dz^{(\mathbf{b})}+\\
 \nonumber+\{\eta_{(\mathbf{0})(\mathbf{0})}+\frac{1}{2}[\frac{1}{2}\epsilon_{(\mathbf{B})}B_{(\mathbf{A}){(\mathbf{B})(\mathbf{C})(\mathbf{D})}}+\\
 \nonumber+\eta^{(\mathbf{M})(\mathbf{N})}A_{(\mathbf{M}){(\mathbf{B})(\mathbf{A})}}A_{(\mathbf{N}){(\mathbf{C})(\mathbf{D})}}].\\
\nonumber.(z^{(\mathbf{B})}\frac{dz^{(\mathbf{A})}}{d\tau}-z^{(\mathbf{A})}\frac{dz^{(\mathbf{B})}}{d\tau})(z^{(\mathbf{C})}\frac{dz^{(\mathbf{D})}}{d\tau}-z^{(\mathbf{D})}\frac{dz^{(\mathbf{C})}}{d\tau})\}d\tau^2,\\
\end{eqnarray}
in which $(a),(b)\neq 0 $.
 \newline
 Defining
\begin{eqnarray}
d\rho^2=\{\eta_{(\mathbf{0})(\mathbf{0})}+\frac{1}{2}[\frac{1}{2}\epsilon_{(\mathbf{B})}B_{(\mathbf{A}){(\mathbf{B})(\mathbf{C})(\mathbf{D})}}+\\
 \nonumber+\eta^{(\mathbf{M})(\mathbf{N})}A_{(\mathbf{M}){(\mathbf{B})(\mathbf{A})}}A_{(\mathbf{N}){(\mathbf{C})(\mathbf{D})}}].\\
\nonumber.(z^{(\mathbf{B})}\frac{dz^{(\mathbf{A})}}{d\tau}-z^{(\mathbf{A})}\frac{dz^{(\mathbf{B})}}{d\tau})(z^{(\mathbf{C})}\frac{dz^{(\mathbf{D})}}{d\tau}-z^{(\mathbf{D})}\frac{dz^{(\mathbf{C})}}{d\tau})\}d\tau^2,\\
\end{eqnarray}
then, (3.1) can be rewritten as
\begin{equation}
ds^2=d\rho^2+\eta_{(\mathbf{a})(\mathbf{b})}dz^{(\mathbf{a})}dz^{(\mathbf{b})}.
\end{equation}
We now write (2.43) as
\begin{eqnarray}
 \nonumber ds^2=\eta_{(\mathbf{A})(\mathbf{B})}dz^{(\mathbf{A})}dz^{(\mathbf{B})}+\\
 \nonumber +\{\frac{1}{2}[\frac{1}{2}\epsilon_{(\mathbf{B})}B_{(\mathbf{A}){(\mathbf{B})(\mathbf{C})(\mathbf{D})}}+\\
 \nonumber +\eta^{(\mathbf{M})(\mathbf{N})}A_{(\mathbf{M}){(\mathbf{B})(\mathbf{A})}}A_{(\mathbf{N}){(\mathbf{C})(\mathbf{D})}}]\}.\\
\nonumber .(z^{(\mathbf{B})}\frac{dz^{(\mathbf{A})}}{ds}-z^{(\mathbf{A})}\frac{dz^{(\mathbf{B})}}{ds})(z^{(\mathbf{C})}\frac{dz^{(\mathbf{D})}}{ds}-z^{(\mathbf{D})}\frac{dz^{(\mathbf{C})}}{ds})ds^2.\\
\end{eqnarray}
\newpage
This can be written in the following form:
\begin{eqnarray}
 \nonumber[1-\frac{1}{2}[\frac{1}{2}\epsilon_{(\mathbf{B})}B_{(\mathbf{A}){(\mathbf{B})(\mathbf{C})(\mathbf{D})}}+\\
 \nonumber+\eta^{(\mathbf{M})(\mathbf{N})}A_{(\mathbf{M}){(\mathbf{B})(\mathbf{A})}}A_{(\mathbf{N}){(\mathbf{C})(\mathbf{D})}}].\\
\nonumber.(z^{(\mathbf{B})}\frac{dz^{(\mathbf{A})}}{ds}-z^{(\mathbf{A})}\frac{dz^{(\mathbf{B})}}{ds})(z^{(\mathbf{C})}\frac{dz^{(\mathbf{D})}}{ds}-z^{(\mathbf{D})}\frac{dz^{(\mathbf{C})}}{ds})]ds^2\\
 \nonumber =\eta_{(\mathbf{A})(\mathbf{B})}dz^{(\mathbf{A})}dz^{(\mathbf{B})}.\\
\end{eqnarray}
We now define the function
\begin{eqnarray}
L^{(\mathbf{B})(\mathbf{A})}=(z^{(\mathbf{B})}\frac{dz^{(\mathbf{A})}}{ds}-z^{(\mathbf{A})}\frac{dz^{(\mathbf{B})}}{ds}),
\end{eqnarray}
which is the classical angular momentum  of a free particle.
\newline
The line element (3.6) can assume the form
\begin{eqnarray}
 \nonumber\{1+\frac{1}{2}[\frac{1}{2}(\epsilon_{(\mathbf{B})}B_{(\mathbf{A}){(\mathbf{B})(\mathbf{C})(\mathbf{D})}}+\\
 \nonumber+\eta^{(\mathbf{M})(\mathbf{N})}A_{(\mathbf{M}){(\mathbf{B})(\mathbf{A})}}A_{(\mathbf{N}){(\mathbf{C})(\mathbf{D})}}].\\
\nonumber.(L^{\mathbf{A})(\mathbf{B})}L^{\mathbf{C})(\mathbf{D})})\}ds^2\\
 \nonumber =(\eta_{(\mathbf{A})(\mathbf{B})}dz^{(\mathbf{A})}dz^{(\mathbf{B})}.\\
\end{eqnarray}
We now define the function
\begin{eqnarray}
\nonumber \exp(-2\sigma)=\{1+\frac{1}{2}[\frac{1}{2}(\epsilon_{(\mathbf{B})}B_{(\mathbf{A}){(\mathbf{B})(\mathbf{C})(\mathbf{D})}}\\
 \nonumber +\eta^{(\mathbf{M})(\mathbf{N})}A_{(\mathbf{M}){(\mathbf{B})(\mathbf{A})}}A_{(\mathbf{N})){(\mathbf{C})(\mathbf{D})}})].\\
\nonumber .L^{(\mathbf{A})(\mathbf{B})}L^{(\mathbf{C})(\mathbf{D})}\},\\
\end{eqnarray}
 then, the line element assumes the form
\begin{equation}
ds^2=\exp(2\sigma)\eta_{(\mathbf{A})(\mathbf{B})}dz^{(\mathbf{A})}dz^{(\mathbf{B})}.
\end{equation}
\newpage
When transformations such as (3.2) are possible, (3.4) is a flat
metric with the time changed and is equivalent to the original
metric. In other words, (3.4) is the original line element of the
curved manifold written in a flat form.
\newline
The metric (3.10) is conformal to a flat manifold as well as a
manifold of constant curvature when the normal coordinate
transformations are well-behaved in the origin and in their
neighborhood. A Riemannian normal transformation and its inverse are
well-behaved in the region in which geodesics are not mixed. Points
in which geodesics approach each other or mix are known as conjugate points of
Jacobi's fields. Jacobi's fields can be used for this purpose.
Although this is an important problem, we do not make further
considerations about the regions in which (3.4) and (3.10) are valid.
\newline
In the next section, we present the Cartan solution for the case
in which curvature is constant. For the Cartan solution to a general
metric, more geometric objects are necessary, such as normal tensors.
This will be presented in section $5$.
\section{Cartan's Solution for Constant Curvature}
  $                      $
 \setcounter{equation}{0}
 $         $
In this section, we present the Cartan solution for constant
curvature. The calculation is very simple and was done in \cite{3},
and reproduced in detail in \cite{4}. Our objective in this section
is only to place the Cartan solution in the forms (3.4) and
(3.10).
\newline
Cartan used the signature $(+,+,+....,+)$ and obtained the
line element
\begin{eqnarray}
 \nonumber
 ds^2=\sum_{k=1}^{n}(\varpi^{\mathbf{k}})^2=\sum_{k=1}^{n}(dv^{\mathbf{k}})^2+\\
 \nonumber -[\frac{|K|{\mathbf{r^2}} -{\mathbf{S^2(r\sqrt{|K|}t)}}}{|K|\mathbf{r^4}}]\sum_{i<j}(v^{\mathbf{i}}dv^{\mathbf{j}}-v^{\mathbf{j}}dv^{\mathbf{i}})^2,\\
\end{eqnarray}
in which when $K>0$
\begin{equation}
  \mathbf{S}=\sin( \sqrt{|K|}t),
\end{equation}
 and when $K<0$
\begin{equation}
  \mathbf{S}=\sinh( \sqrt{|K|}t).
\end{equation}
 We write (4.1) in the form (3.1)
\begin{eqnarray}
 \nonumber
 ds^2=\sum_{k=1}^{n}(\varpi^{\mathbf{k}})^2=\sum_{k=1}^{n}(dv^{\mathbf{k}})^2\\
 \nonumber -[\frac{|K|{\mathbf{r^2}} -{\mathbf{S^2(r\sqrt{|K|}t)}}}{|K|\mathbf{r^4}}]\sum_{i<j}(v^{\mathbf{i}}\frac{dv^{\mathbf{j}}}{d\tau}-v^{\mathbf{j}}\frac{dv^{\mathbf{i}}}{d\tau})^2d\tau^2.\\
\end{eqnarray}
 Consider the following function:
\begin{equation}
l^{\mathbf{ij}}=\sum_{i<j}(v^{\mathbf{i}}\frac{dv^{\mathbf{j}}}{d\tau}-v^{\mathbf{j}}\frac{dv^{\mathbf{i}}}{d\tau})^2.
\end{equation}
\newline
Using (4.5) in (4.4), we have
\begin{eqnarray}
\nonumber
 ds^2=\sum_{k=1}^{n}(\varpi^{\mathbf{k}})^2=\sum_{k=1}^{n}(dv^{\mathbf{k}})^2\\
 \nonumber -[\frac{|K|{\mathbf{r^2}} -{\mathbf{S^2(r\sqrt{|K|}t)}}}{|K|\mathbf{r^4}}]\sum_{i<j}(l^{\mathbf{ij}})^2d\tau^2.\\
\end{eqnarray}
Sometimes, we can put $ dv^1=d\tau $. In this case,
(4.6) can be written in the form
\begin{eqnarray}
 \nonumber ds^2=\sum_{k=1}^{n}(\varpi^{\mathbf{k}})^2=\sum_{k=2}^{n}(dv^{\mathbf{k}})^2\\
 \nonumber +\{1 -[\frac{|K|{\mathbf{r^2}} -{\mathbf{S^2(r\sqrt{|K|}t)}}}{|K|\mathbf{r^4}}]\sum_{i<j}(l^{\mathbf{ij}})^2\}d\tau^2.\\
\end{eqnarray}
Defining
\begin{equation}
 d\rho^2=\{1-[\frac{|K|{\mathbf{r^2}}-{\mathbf{S^2(r\sqrt{|K|}t)}}}{|K|\mathbf{r^4}}]\sum_{i<j}(l^{\mathbf{ij}})^2\}d\tau^2\\,
\end{equation}
and placing it in (4.6), we obtain
\begin{equation}
ds^2=d\rho^2 +\sum_{k=2}^{n}(dv^{\mathbf{k}})^2,
\end{equation}
in which (4.9) has the same form as (3.4).
\newline
We now write (4.1) in the form (3.10). For this, we change (4.1) as
follows:
\begin{eqnarray}
 \nonumber
 ds^2=\sum_{k=1}^{n}(\varpi^{\mathbf{k}})^2=\sum_{k=1}^{n}(dv^{\mathbf{k}})^2+\\
 \nonumber -[\frac{|K|{\mathbf{r^2}} -{\mathbf{S^2(r\sqrt{|K|}t)}}}{|K|\mathbf{r^4}}]\sum_{i<j}(v^{\mathbf{i}}\frac{dv^{\mathbf{j}}}{ds}-v^{\mathbf{j}}\frac{dv^{\mathbf{i}}}{ds})^2ds^2.\\
\end{eqnarray}
We see that (4.10) has the form of (3.5).
\newline
Defining
\begin{eqnarray}
L^{(\mathbf{i})(\mathbf{j})}=(z^{(\mathbf{i})}\frac{dz^{(\mathbf{j})}}{ds}-z^{(\mathbf{j})}\frac{dz^{(\mathbf{i})}}{ds}).
\end{eqnarray}
and placing (4.11) in (4.10), we obtain
 \begin{eqnarray}
 \nonumber
 ds^2=\sum_{k=1}^{n}(\varpi^{\mathbf{k}})^2=\sum_{k=1}^{n}(dv^{\mathbf{k}})^2+\\
 \nonumber -[\frac{|K|{\mathbf{r^2}} -{\mathbf{S^2(r\sqrt{|K|}t)}}}{|K|\mathbf{r^4}}]\sum_{i<j}(L^{(\mathbf{i})(\mathbf{j})})^2ds^2,\\
\end{eqnarray}
which is equivalent to
\begin{eqnarray}
\{1+[\frac{|K|{\mathbf{r^2}}
-{\mathbf{S^2(r\sqrt{|K|}t)}}}{|K|\mathbf{r^4}}]\sum_{i<j}(L^{(\mathbf{i})(\mathbf{j})})^2\}ds^2=\sum_{k=1}^{n}(dv^{\mathbf{k}})^2.
\end{eqnarray}
We now define
\begin{eqnarray}
\exp(-2\sigma)=\{1+[\frac{|K|{\mathbf{r^2}}
-{\mathbf{S^2(r\sqrt{|K|}t)}}}{|K|\mathbf{r^4}}]\sum_{i<j}(L^{(\mathbf{i})(\mathbf{j})})^2\}.
\end{eqnarray}
Placing in (4.13), we obtain
\begin{equation}
ds^2=\exp(2\sigma)\sum_{k=1}^{n}(dv^{\mathbf{k}})^2.
\end{equation}
We could have all equations in this section on a vielbein basis and
the results would be the same. This will be made at the end of the
next section for the general solution.
\newline
We rewrite (4.15) as
\begin{eqnarray}
ds^2=\{1+[\frac{|K|{\mathbf{r^2}}
-{\mathbf{S^2(r\sqrt{|K|}t)}}}{|K|\mathbf{r^4}}]\sum_{i<j}(\eta_{\mathbf{i}\mathbf{j}}L^{(\mathbf{i})(\mathbf{j})})^2\}^{-1}dv^{\mathbf{l}}dv^{\mathbf{k}}\eta_{\mathbf{l}\mathbf{k}},
\end{eqnarray}
in which  $\eta_{\mathbf{j}\mathbf{k}}$ is a generic flat metric.
\newline
Through a coordinate transformation, we can put (4.16) in the well-known
form
\begin{eqnarray}
ds^2=\{1+\frac{K\Omega^{\mathbf{j}}\Omega^{\mathbf{k}}\eta_{\mathbf{j}{\mathbf{k}}}}{4}\}^{-2}d\Omega^{\mathbf{j}}d\Omega^{\mathbf{k}}\eta_{\mathbf{j}{\mathbf{k}}}.
\end{eqnarray}
 But (4.17) is conformal to a flat metric. As
(4.16) and (4.17) are equivalent, (4.16) is also
conformal to a flat metric.
\newline
In the next section, we present some geometric objects in detail, such as
normal tensors. This is necessary for the Cartan solution of
a general metric.
\newline
\section{Normal Tensors}
  $                      $
 \setcounter{equation}{0}
 $         $
In this section, Taylor's expansion will be
built in the origin of normal coordinates for the metric tensor components.
Normal tensors are very important for this. For the covariant
derivative, we use the notation (;).
\newline
Consider the line element
\begin{equation}
 ds^2= G_{\Lambda\Pi}du^{\Lambda}du^{\Pi}.
\end{equation}
Its expansion has the general form in the origin of a normal coordinate
\begin{eqnarray}
 \nonumber ds^2=G_{\lambda\pi}du^{\lambda}du^{\pi}=G_{\lambda\pi}(0)+\frac{\partial{G_{\lambda\pi}}}{\partial{u^{\mu}}}v^{\mu}t\\
\nonumber+\frac{1}{2}\frac{\partial^2G_{\lambda\pi}}{\partial{u^{\mu}}\partial{u^{\nu}}}v^{\mu}v^{\nu}t^2+.......,\\
\end{eqnarray}
in which the derivatives are calculated at $ u^{\pi}=0 $.
\newline
Some results are found in \cite{6}, \cite{7}, but these are not generally simple.
Our results are simpler because they are more specific.
\newline
Consider the covariant derivative of $G_{\lambda\pi}$ in a normal
coordinate system.
\newline
For a pseudo-Riemannian space, we have
\begin{equation}
 G_{\lambda\pi};_{\mu}=0.
\end{equation}
From (5.3), we obtain
\begin{equation}
 \frac{\partial{G_{\lambda\pi}}}{\partial{u^{\mu}}}=C^{\rho}_{\mu\lambda}G_{\rho\pi}
+C^{\rho}_{\mu\pi}G_{\lambda\rho},
\end{equation}
in which
\begin{equation}
C^{\rho}_{\mu\lambda}(0)=0,
\end{equation}
and
\begin{equation}
 \frac{\partial{G_{\lambda\pi}}}{\partial{u^{\mu}}}(0)=0,
\end{equation}
in the origin.
\newline
In the  limit  $u=0$, the  partial derivatives of (5.4) supply all
derivative terms for the expansion (5.2). Each partial derivative of
$C^{\rho}_{\mu\lambda}$, calculated in the origin, is a new tensor.
These new tensors are denominated normal tensors. We designate the
following representation for these tensors:
\begin{equation}
D^{\rho}_{\mu\lambda\alpha\beta....\gamma}=
\frac{\partial^nC^{\rho}_{\mu\lambda}}{\partial{u^{\alpha}}\partial{u^{\beta}}...\partial{u^{\gamma}}}(0).
\end{equation}
We conclude from (5.7) that normal tensors are symmetric in the
first pair of inferior indices and have a complete symmetry
among other inferior indices.
\newpage
It is simple to show that
\begin{equation}
S(D^{\rho}_{\mu\lambda\alpha\beta....\gamma})=0,
\end{equation}
in which S designates the sum of different normal tensor components.
\newline
With (5.4),(5.5), (5.6), (5.7) and (5.8), we can calculate all terms
of the expansion (5.2).
\newline
Deriving (5.5), calculating the limit and using (5.7), we have
\begin{equation}
 \frac{\partial^2{G_{\lambda\pi}}}{\partial{u^{\mu}}{u^{\nu}}}=G_{\lambda\rho}D^{\rho}_{\mu\pi\nu}
+G_{\pi\rho}D^{\rho}_{\mu\lambda\nu}.
\end{equation}
There is more than one way of associating the curvature tensor with
normal tensors. Next, we present the simplest way we
know.
\newline
Let us define the components of
the Riemannian curvature tensor in normal coordinates:
\begin{equation}
R^{\rho}_{\mu\lambda\nu}=\frac{\partial(C^{\rho}_{\mu\lambda})}{\partial{u^{\nu}}}-\frac{\partial(C^{\rho}_{\mu\nu})}{\partial{u^{\lambda}}}
+C^{\sigma}_{\mu\lambda}C^{\rho}_{\sigma\nu}-C^{\sigma}_{\mu\nu}C^{\rho}_{\sigma\lambda}.
\end{equation}
The limit of (5.10) is
\begin{equation}
R^{\rho}_{\mu\lambda\nu}=D^{\rho}_{\mu\lambda\nu}-D^{\rho}_{\mu\nu\lambda},
\end{equation}
in which we have used (5.5) and (5.7).
\newline
Using (5.7), (5.8),  (5.11) and the symmetries of the Riemannian
curvature tensor, we can show that
\begin{equation}
D^{\rho}_{\mu\lambda\nu}=\frac{1}{3}(R^{\rho}_{\mu\lambda\nu}+R^{\rho}_{\lambda\mu\nu}).
\end{equation}
Using (5.9) and (5.12), we obtain
\begin{equation}
 \frac{\partial^2{G_{\alpha\beta}}}{\partial{u^{\gamma}}{u^{\delta}}}u^{\gamma}u^{\delta}=\frac{2}{3}R_{\alpha\gamma\beta\delta}u^{\gamma}u^{\delta}.
\end{equation}
Through a similar procedure, but a tedious calculation, we obtain
\begin{equation}
 \frac{\partial^3{G_{\alpha\beta}}}{\partial{u^{\mu}}{u^{\nu}}{u^{\sigma}}}u^{\mu}u^{\nu}u^{\sigma}
 =R_{\alpha\mu\beta\nu;\sigma}u^{\mu}u^{\nu}u^{\sigma}.
\end{equation}
Fourth-order derivatives for metric tensors are easy, but very
long.
\newline
So we will not present them here. We can now conclude Taylor's
\newline
expansion of the metric tensor. First, we rewrite
\begin{eqnarray}
 \nonumber G_{\lambda\pi}=G_{\lambda\pi}(0)\\
 \nonumber +\frac{1}{2}\frac{\partial^2G_{\lambda\pi}}{\partial{u^{\mu}}\partial{u^{\nu}}}v^{\mu}v^{\nu}t^2\\
 \nonumber \frac{1}{6}\frac{\partial^3{G_{\alpha\beta}}}{\partial{u^{\mu}}{u^{\nu}}{v^{\sigma}}}v^{\mu}v^{\nu}v^{\sigma}t^3+...,\\
\end{eqnarray}
 Now we place (5.13) and (5.14) in (5.15), obtaining
 \begin{eqnarray}
 \nonumber G_{\lambda\pi}du^{\alpha}du^{\beta}=G_{\alpha\beta}(0)du^{\alpha}du^{\beta}+\\
 \nonumber +\frac{1}{3}[R_{\alpha\gamma\beta\delta}t^2+\\
 \nonumber+ \frac{1}{2}v^{\sigma}R_{\alpha\mu\beta\nu;\sigma}t^3+...]v^{\gamma}v^{\delta}du^{\alpha}du^{\beta},.\\
\end{eqnarray}
Using the symmetries of the curvature tensor, we have the following
expansion:
\begin{eqnarray}
 \nonumber G_{\lambda\pi}du^{\alpha}du^{\beta}=G_{\alpha\beta}(0)du^{\alpha}du^{\beta}+\\
 \nonumber +\frac{1}{12}[R_{\alpha\gamma\beta\delta}t^2+\\
 \nonumber+ \frac{1}{2}v^{\sigma}R_{\alpha\gamma\beta\delta;\sigma}t^3+...][v^{\gamma}du^{\alpha}-v^{\alpha}du^{\gamma}][v^{\beta}du^{\delta}-v^{\delta}du^{\beta}].\\
\end{eqnarray}
On the hyper-surface $t=1$, we have $dt=0$ and
\begin{eqnarray}
 \nonumber G_{\lambda\pi}du^{\alpha}du^{\beta}=G_{\alpha\beta}(0)dv^{\alpha}dv^{\beta}+\\
 \nonumber +\frac{1}{12}[R_{\alpha\gamma\beta\delta}+\\
 \nonumber+ \frac{1}{2}v^{\sigma}R_{\alpha\gamma\beta\delta;\sigma}][v^{\gamma}dv^{\alpha}-v^{\alpha}dv^{\gamma}][v^{\beta}dv^{\delta}-v^{\delta}dv^{\beta}],\\
\end{eqnarray}
which is the same as Cartan's result, although through a different way.
\newline
It is always possible to place a flat metric in a diagonal form.
\newpage
This is the case of a metric at the origin of normal coordinates.
In this case, we have
\begin{equation}
E_{\Lambda}^{(\mathbf{A})}(0)=\delta_{\Lambda}^{(\mathbf{A})}.
\end{equation}
We now present Taylor's expansion of
$E_{\Lambda}^{(\mathbf{A})}$ at the origin of a normal
coordinate
\begin{eqnarray}
\nonumber E_{\Lambda}^{(\mathbf{A})}(u)=\delta_{\Lambda}^{(\mathbf{A})}+\\
 \nonumber+\frac{\partial(E_{\Lambda}^{(\mathbf{A})})}{\partial(u^{\alpha})}du^{\alpha}+...\\
\end{eqnarray}
\newline
Multiplying (5.18) by the vielbein components and their
inverse, using (5.19) and (5.20), we obtain
\begin{eqnarray}
\nonumber ds^2=\eta_{(\mathbf{A})(\mathbf{B})}dz^{(\mathbf{A})}dz^{(\mathbf{B})}+\\
 \nonumber+\frac{1}{12}[R_{(\mathbf{A}){(\mathbf{B})(\mathbf{C})(\mathbf{D})}}+\\
 \nonumber+\frac{1}{2}z^{(\mathbf{M})}R_{(\mathbf{A}){(\mathbf{B})(\mathbf{C})(\mathbf{D}),(\mathbf{M})}}].\\
\nonumber.(z^{(\mathbf{B})}dz^{(\mathbf{A})}-z^{(\mathbf{A})}dz^{(\mathbf{B})})(z^{(\mathbf{C})}dz^{(\mathbf{D})}-z^{(\mathbf{D})}dz^{(\mathbf{C})}),\\
\end{eqnarray}
in which the calculation was made on the hyper-surface $t=1$ and
$dt=0$.
\newline
Note that the expansion given by (5.21) is an approximated solution
of (2.43). Using a perturbation method, Cartan  first solved  the
equations (2.28) and (2.34) and then placed each solution in (2.43).
\newline
From an appropriate coordinate transformation, we can put a metric
of an n-dimensional manifold of constant curvature in the form
\begin{eqnarray}
ds'^2=\{1+\frac{K\Omega^{\mathbf{\alpha}}\Omega^{\beta}\eta_{\alpha\beta}}{4}\}^{-2}d\Omega^{\rho}d\Omega^{\sigma}\eta_{\varrho\sigma}.
\end{eqnarray}
We can build an expression between $ d\Omega^{\rho} $ and
$dz^{(\mathbf{A})}$, as follows:
\begin{eqnarray}
d\Omega^{\Lambda}=E_{(\mathbf{A})}^{\Lambda}dz^{(\mathbf{A})}.
\end{eqnarray}
Then, placing (5.23) in (3.10), we have
\begin{eqnarray}
\nonumber ds^2=\{1+\frac{1}{2}[\frac{1}{2}(\epsilon_{\beta}B_{\alpha{\beta\gamma\delta}})\\
 \nonumber +\eta^{\rho\sigma}A_{\rho{\alpha\beta}}A_{\sigma{\gamma\delta)}})].\\
\nonumber .L^{\alpha\beta}L^{\gamma\delta}\}^{-1}\eta_{\alpha\beta}d\Omega^{\alpha}d\Omega^{\beta}.\\
\end{eqnarray}
Because (5.22) and (5.24) are conformal to a flat manifold, there is
a conformal transformation between them, with a conformal factor,
$(\exp2\psi)$. Then
\begin{eqnarray}
g'_{\alpha\beta}=(\exp2\psi)g_{\alpha\beta}.
\end{eqnarray}
More specifically,
\begin{eqnarray}
\nonumber\{1+\frac{1}{2}[\frac{1}{2}(\epsilon_{\beta}B_{\alpha{\beta\gamma\delta}})+\\
\nonumber
+\eta^{\rho\sigma}A_{\rho{\alpha\beta}}A_{\sigma{\gamma\delta)}})]L^{\alpha\beta}L^{\gamma\delta}\}=\\
 \nonumber
 =(\exp2\psi)\{1+\frac{K\Omega^{\mathbf{\alpha}}\Omega^{\beta}\eta_{\alpha\beta}}{4}\}^{2}.\\
 \end{eqnarray}
\newline
This is an important result with some consequences, as we will see later.
\newline
If we make the Lie derivative of (5.25) in relation to $\xi$, we
obtain
\begin{eqnarray}
L_{\xi}g'=(2\xi(\psi)g+L_{\xi}g)(\exp2\psi).
\end{eqnarray}
We now consider the condition
\begin{eqnarray}
2\xi(\psi)g+L_{\xi}g=0,
\end{eqnarray}
which implies
\begin{eqnarray}
L_{\xi}g'=0,
\end{eqnarray}
which is a definition of a Killing vector. The solution
to (5.29) is well known, obtaining each of the Killing vectors $\xi$. We conclude that
each $\xi$ is a Killing vector in (5.22) and a conformal Killing vector
in (5.24). The equations (5.28) and (5.29) show how a Killing vector in
(5.22) will be a conformal Killing vector in (5.24). Notice that a
pseudo-Riemannian metric can be put in the form (5.24) by a Riemannian
normal transformation. In (5.24), we have a conformal Killing vector, which
is a Killing vector in (5.22).
\newpage
\section{Embedding Manifolds of Constant Curvature in Flat Manifolds}
 $                       $
 \setcounter{equation}{0}
 $         $
There are many procedures for defining or introducing functions, fields
and geometric objects in an n-sphere. From a different point of view,
Dirac \cite{8} embedded the De Sitter space in a flat five-dimensional
manifold. The author considered functions and fields as residing in a flat five-dimensional
manifold and constructed a procedure to project these functions and fields in the De Sitter space.
The Dirac procedure implies a need for the quantum momentum and quantum
angular momentum postulates. Other authors have used Dirac's ideas or
variants of these ideas, as in \cite{9}, where an n+1-dimensional stereographic projection
is used in a way in which the quantum angular momentum remains in the n-sphere.
This does not happen in our approach, because the quantum angular momentum resides on
an n+1-dimensional pseudo-sphere, as we shall see.
\newline
In this section, we embed the n-dimensional manifold (5.22) in a flat n+1-dimensional
manifold, obtaining the quantum angular momentum of a free particle as the geometric result.
\newline
We now consider a manifold (5.22) designated by S embedded in a flat
n+1-dimensional manifold. The following constraint is obeyed
\cite{10}:
\begin{eqnarray}
\eta_{\alpha\beta}x^{\alpha}x^{\beta}=K=\epsilon\frac{1}{R^2},
\end{eqnarray}
in which K is the scalar curvature of the n-dimensional manifold
(5.22) and $\alpha, \beta = ( 1, 2,...,n+1 )$ and $ \epsilon=(+1, -1).$
\newline
It is suitable to use a local basis
$X_{\beta}=\frac{\partial}{\partial(x^{\beta})}.$
\newline
We consider a constant vector $\textbf{C}$ in a flat n+1-dimensional
manifold given by
\begin{eqnarray}
\eta_{\alpha\beta}C^{\alpha}X^{\beta}=\eta^{\alpha\beta}C_{\alpha}X_{\beta}=C,
\end{eqnarray}
in which  $C^{\alpha}$ are constant and $\textbf{N}$ is an
orthonormal vector to S. We use the symbol $<,>$ for the inner product
in the flat n+1-dimensional manifold and $<,>'$ for S.
\newline
A constant vector $\textbf{C}$ can be decomposed into two parts,
one on S and the other outside, as follows:
\begin{eqnarray}
C=\bar{C}+<C,N>N.
\end{eqnarray}
From the definition of $\textbf{N}$ and (6.1), we obtain
\begin{eqnarray}
N^{\alpha}=\frac{x^{\alpha}}{R}
\end{eqnarray}
Let us construct the covariant derivative of $\textbf{C}$. We have
a local basis and a diagonal, unitary metric tensor, so that the
Christoffel symbols are null. Thus, the covariant derivative of
$\textbf{C}$ in the $\textbf{Y}$ direction is given by
\begin{eqnarray}
\nabla_{Y}C=0.
\end{eqnarray}
It is easy to show that
\begin{eqnarray}
\nabla_{Y}N=\frac{Y}{R}.
\end{eqnarray}
The Lie derivative of the metric tensor in S is given by \cite{10}:
\begin{eqnarray}
L_{\bar{U}}g'=2\lambda_{U}g',
\end{eqnarray}
in which $\textbf{U}$ is a constant vector in the flat manifold and
$\lambda_{U}$ is the characteristic function. For S, the
characteristic function is given by
\begin{eqnarray}
\lambda_{U}=-\frac{1}{R}<U,N>.
\end{eqnarray}
Placing (6.8) in (6.7), we have
\begin{eqnarray}
L_{\bar{U}}g'=-2\frac{1}{R}<U,N>g'.
\end{eqnarray}
In the region of S in which $<U,N>$ is not vanished, $ \bar{U} $ is a
conformal Killing vector and in the region in which $<U,N>$ is vanished,
 $\bar{U} $ is a Killing vector.
\newline
We now consider another constant vector $\textbf{V}$ in the flat
space. The Lie derivative of its projection on S is given by
\begin{eqnarray}
L_{\bar{V}}g'=-2\frac{1}{R}<V,N>g'.
\end{eqnarray}
As we consider a local basis and constant vectors $\textbf{U}$ and
$\textbf{V}$, the commutator is given by
\begin{eqnarray}
[U,V]=0.
\end{eqnarray}
Then,
\begin{eqnarray}
L_{[\bar{U},\bar{V}]}g'=-2\frac{1}{R}<[U,V],N>g'=0.
\end{eqnarray}
\newline
Regardless of $ \bar{U}$ and $\bar{V}$ being Killing or conformal
Killing vectors, their commutator is a Killing object. We will now
show that the commutator $[\bar{U},\bar{V}]$ is
proportional to the quantum angular momentum of a particle.
\newline
Using (6.3) in the commutator of elements of the basis,
we obtain
\begin{eqnarray}
\nonumber [\bar{U},\bar{V}]=\\
\nonumber =U^{\alpha}V^{\beta}[X_\alpha-<X_\alpha,N>N,X_\beta-<X_\beta,N>N]=\\
\nonumber =U^{\alpha}V^{\beta}[\bar{X}_\alpha,\bar{X}_\beta].\\
\end{eqnarray}
We now calculate the commutator of elements of the basis by parts.
\newline
We have by simple calculation
\begin{eqnarray}
<X_\alpha,N>=\frac{1}{R}\eta_{\alpha\beta}x^{\beta}.
\end{eqnarray}
Placing (6.14) in (6.13), we obtain
\begin{eqnarray}
\nonumber [\bar{X}_\alpha,\bar{X}_\beta]=\\
\nonumber =[X_\alpha,X_\beta]-[X_\alpha,\frac{1}{R}\eta_{\beta\sigma}x^{\sigma}N]+[X_\beta,\frac{1}{R}\eta_{\alpha\sigma}x^{\sigma}N]+\\
\nonumber +\frac{1}{R^2}[\eta_{\alpha\sigma}x^{\sigma}N, \eta_{\beta\sigma}x^{\sigma}N].\\
\end{eqnarray}
On a local basis, we have
\begin{eqnarray}
[X_\alpha,X_\beta]=0,
\end{eqnarray}
\begin{eqnarray}
[\eta_{\alpha\sigma}x^{\sigma}N, \eta_{\beta\sigma}x^{\sigma}N]=0.
\end{eqnarray}
Placing (6.16) and (6.17) in (6.15), we obtain
\begin{eqnarray}
\nonumber [\bar{X}_\alpha,\bar{X}_\beta]=\\
\nonumber =
\frac{1}{R^2}(\eta_{\alpha\sigma}x^{\sigma}\frac{\partial}{\partial(x^{\beta})}-\eta_{\beta\sigma}x^{\sigma}\frac{\partial}{\partial(x^{\alpha})})\\
\nonumber =
\frac{1}{R^2}(x_\alpha\frac{\partial}{\partial(x^{\beta})}-x_\beta\frac{\partial}{\partial(x^{\alpha})})\\
\nonumber=-i\frac{1}{\hbar}\frac{1}{R^2}L_{\alpha\beta}.\\
\end{eqnarray}
Multiplying $ L_{\alpha\beta}$ by a vielbein basis, we obtain
\begin{eqnarray}
\nonumber
L_{(\mathbf{A})(\mathbf{B})}=\\
\nonumber
=(i\hbar)(R^{2})R_{(\mathbf{A})(\mathbf{B})(\mathbf{C})(\mathbf{D})}x^{(\mathbf{D})}\eta^{(\mathbf{C})(\mathbf{M})}\frac{\partial}{\partial(x^{\mathbf{M})}}.\\
\end{eqnarray}
in which
\begin{eqnarray}
 \hat{p}_{(\mathbf{M})}=(i\hbar)\frac{\partial}{\partial(x^{\mathbf{M})}}
\end{eqnarray}
is the quantum momentum operator of a particle and
\begin{eqnarray}
\nonumber
R_{(\mathbf{A})(\mathbf{B})(\mathbf{C})(\mathbf{D})}=\\
\nonumber
=\frac{1}{R^2}[\eta_{(\mathbf{A})(\mathbf{D})}\eta_{(\mathbf{B})(\mathbf{C})}-\eta_{(\mathbf{A})(\mathbf{C})}\eta_{(\mathbf{B})(\mathbf{D})}]\\
\end{eqnarray}
is the curvature of S on the vielbein basis and
$ \eta_{(\mathbf{A})(\mathbf{C})}$ is diagonal.
\newline
We have used the embedding of an n-dimensional manifold S in a
flat n+1-dimensional manifold only to obtain the quantum
angular momentum operator of a particle.
\newline
We can rewrite (6.19) as follows:
\begin{eqnarray}
\nonumber
L_{(\mathbf{A})(\mathbf{B})}=\\
\nonumber
=(i\hbar)[\eta_{(\mathbf{A})(\mathbf{D})}\eta_{(\mathbf{B})(\mathbf{C})}-\eta_{(\mathbf{A})(\mathbf{C})}\eta_{(\mathbf{B})(\mathbf{D})}].\\
\nonumber
.x_{(\mathbf{D})}\eta^{(\mathbf{C})(\mathbf{M})}\frac{\partial}{\partial(x^{\mathbf{M})}}.\\
\end{eqnarray}
\newline
Racah \cite{11} has shown that the Casimir operators of semisimple
Lie groups can be constructed from the quantum angular momentum (6.19).
Each multiplet of a semisimple Lie group can be uniquely characterized by
the eigenvalues of the Casimir operators.
\newline
Although we have built the quantum angular momentum from classical
geometric considerations, we can write the usual expression for an
eigenstate of Casimir operator as follows:
\begin{eqnarray}
\hat{C}\mid...>=C\mid...>.
\end{eqnarray}
Next, we calculate the Lie derivative of the so(p,n-p) algebra.
\newline
For the Lie group SO(p,q), we choose the signature
\newline
$(p,q)=(p,n-p)=(-,-,-,...-,+,+,..+)$,
\newline
with the algebra
\begin{eqnarray}
\nonumber
[L_{(\mathbf{A})(\mathbf{B})},L_{(\mathbf{C})(\mathbf{D})}]=-i(\eta_{(\mathbf{A})(\mathbf{C})}L_{(\mathbf{B})(\mathbf{D})}+\eta_{(\mathbf{A})(\mathbf{D})}L_{(\mathbf{C})(\mathbf{B})}\\
\nonumber +\eta_{(\mathbf{B})(\mathbf{C})}L_{(\mathbf{D})(\mathbf{A})}+\eta_{(\mathbf{B})(\mathbf{D})}L_{(\mathbf{A})(\mathbf{C})}).\\
\end{eqnarray}
Considering the Lie derivative
\begin{eqnarray}
\nonumber \textbf{L}_{[L_{(\mathbf{A})(\mathbf{B})},L_{(\mathbf{C})(\mathbf{D})}]}g'=\\
\nonumber
=-R^{4}<[[X_{(\mathbf{A})},X_{(\mathbf{B})}],[X_{(\mathbf{C})},X_{(\mathbf{D})}]],N>g'=0.\\
\end{eqnarray}
The vielbein for orthogonal Cartesian coordinates is
given by
\begin{equation}
E_{\Lambda}^{(\mathbf{A})}=\delta_{\Lambda}^{(\mathbf{A})}.
\end{equation}
We then have
\begin{eqnarray}
\nonumber[X_{(\mathbf{A})},X_{(\mathbf{B})}]=[X_{\alpha},X_{\beta}]=0.\\
\end{eqnarray}
\newline
Notice that $g'$ in S is form-invariant in relation to the
Killing vector $\xi$ \cite{12} and the algebra of
SO(p,n-p). We conclude that the algebra of SO(p,n-p) is a Killing
object. The same is true for the algebra of the Lie group SO(n),
in which we could choose the signature $(+,+,+...,+,+)$ for SO(n).
\newline
The constraint (6.1) is invariant for many classical groups.
For such groups, it is possible to build operators from the
combination of the quantum angular momentum operators, which are
Killing objects in relation to $´g'$. Therefore, the metric is
form-invariant in relation to this algebra. Some of these groups
are discussed in \cite{12} and Cartan's list of irreducible Riemannian
globally symmetric spaces in \cite{5}.
\newline
Notice that we start from a normal coordinate transformation. In other
words, in the region in which the transformation (2.4) is well-behaved,
we can build (3.10) and we have (5.22) by a conformal transformation,
which was essential to obtaining the quantum angular momentum operator
from geometry.
\newline
The coordinates $ X^{\Pi}$ of the orthonormal Cartesian coordinate
frame, which appear in (6.1), reside in the flat n+1-dimensional manifold
and describe an n-dimensional pseudo-sphere. A free classical particle,
which resides on an n-dimensional pseudo-sphere, obeys the same classical
angular momentum expression (3.7), which appears in (3.8)and (3.9),
obtained in a different context, and obeys
$ L^{\lambda\alpha}\subset$ S, with $\alpha,\lambda=(1,2,....,n)$.
This is not true for the quantum angular momentum $L_{(\mathbf{A})(\mathbf{B})}$
given by (6.19), because in
$R_{(\mathbf{A})(\mathbf{B})(\mathbf{C})(\mathbf{D})}$ we have
$(A),(B),(C),(D)=(1,2,....,n,n+1)$. This is not easily or always noticed.
We now present a detailed explanation. To avoid a confusing notation,
we rewrite (6.1) as follows \cite{13}:
\begin{eqnarray}
\bar{G}_{\Pi\Lambda}X^{\Pi}X^{\Lambda}= K\eta_{\alpha\beta}x^{\alpha}x^{\beta}+z^{2}=1,
\end{eqnarray}
in which K is the scalar curvature of the n-dimensional manifold (5.22),
$\alpha, \beta = ( 1, 2,...,n)$ and $\Lambda,\Pi = ( 1, 2,...,n+1 ).$
\newline
The line element of the flat n+1-dimensional manifold is given by
\begin{eqnarray}
 ds^{2}=\tilde{G}_{\Pi\Lambda}dX^{\Pi}dX^{\Lambda}=\eta_{\alpha\beta}dx^{\alpha}dx^{\beta}+K^{-1}dz^{2},
\end{eqnarray}
After a simple calculation, we have the metric of the n-dimensional pseudo-sphere
\begin{eqnarray}
 g_{\mu\nu}=\eta_{\mu\nu}+\frac{K}{(1- K\eta_{\rho\sigma}x^{\rho}x^{\sigma})}\eta_{\mu\alpha}x^{\alpha}\eta_{\nu\beta}x^{\beta},
\end{eqnarray}
in which we have chosen a diagonal metric with each $ \eta_{\mu\nu}=\epsilon_{\mu}\delta_{\mu\nu}$
as plus or minus Kronecker's  delta function. After a long, but simple consideration of (6.29)
and (6.30), we put the line element (6.29) in the form
\begin{eqnarray}
 ds^{2}=\frac{d\bar{s}^{2}}{(1-Kx^{2})[1+K\sum_{\alpha>\lambda}\eta_{\lambda\rho}\eta_{\alpha\sigma}L^{\lambda\alpha}L^{\rho\sigma}\}]}
\end{eqnarray}
for which, in (6.31), we have $\alpha=(2,3,......,n),$
 \begin{eqnarray}
 x^{2}=\eta_{\rho\sigma}x^{\rho}x^{\sigma},
\end{eqnarray}
\begin{eqnarray}
L^{\lambda\alpha}=x^{\lambda}\frac{dx^{\alpha}}{ds}-x^{\alpha}\frac{dx^{\lambda}}{ds},
\end{eqnarray}
and
\begin{eqnarray}
 d\bar{s}^{2}=\eta_{\rho\sigma}dx^{\rho}dx^{\sigma}.
\end{eqnarray}
The line element of the n-dimensional pseudo-sphere given by (6.31) is in the
conformal form and a particle in free motion described by the classical angular momentum
(6.33) resides on this n-dimensional pseudo-sphere. We are going to build Riemann's
tensor components for the metric (6.30). Placing (6.30) in Christoffel's symbols,
after some calculation, we have
\begin{equation}
\Gamma_{\mu \nu}^{\eta}=Kx^{\eta} g_{\mu\nu}.
\end{equation}
 Now consider Riemann's tensor components:
\begin{equation}
R^{\alpha}{}_{\mu \sigma \nu }=\partial_{\nu} \Gamma_{\mu \sigma
}^{\alpha}-\partial_{\sigma}\Gamma_{\mu \nu}^{\alpha}+\Gamma_{\mu
\sigma}^{\eta}\Gamma_{n \nu}^{\alpha}-\Gamma_{\mu \nu}^{\eta}\Gamma
_{\sigma \eta}^{\alpha}.
\end{equation}
Placing (6.35) in (6.36), we obtain
 \begin{equation}
 R_{\lambda \pi \sigma
 \rho}=K(g_{\lambda\rho}g_{\pi\sigma}-g_{\lambda\sigma}g_{\pi\rho}).
\end{equation}
In the origin, $ x^{\mu}=0 $, Riemann's tensor components are given by
\begin{equation}
 R_{\lambda \pi \sigma
 \rho}=K(\eta_{\lambda\rho}\eta_{\pi\sigma}-\eta_{\lambda\sigma}\eta_{\pi\rho}).
\end{equation}
Notice that an n-dimensional pseudo-sphere is an isotropic, homogeneous manifold.
All points on the n-dimensional pseudo-sphere are equivalent. Through continuous coordinate
transformations, we get from (6.38) to (6.37). Now consider the quantum angular momentum
given by (6.19). Multiplying $R_{(\mathbf{A})(\mathbf{B})(\mathbf{C})(\mathbf{D})}$
by an appropriate vielbein basis
$ E_{\Lambda}^{(\mathbf{A})}E_{\Pi}^{(\mathbf{B})}E_{\Theta}^{(\mathbf{C})}E_{\Xi}^{(\mathbf{D})},$
and using
\begin{equation}
G_{\Lambda\Pi}=E_{\Lambda}^{(\mathbf{A})}E_{\Pi}^{(\mathbf{B})}\eta_{(\mathbf{A})(\mathbf{B})},
\end{equation}
we have
\begin{equation}
R_{\Lambda \Pi \Omega
\Theta}=K(G_{\Lambda\Theta}G_{\Pi\Omega}-G_{\Lambda\Omega}G_{\Pi\Theta}),
\end{equation}
in which $(\Lambda,\Pi,\Theta,\Omega)=(1,2,3,....,n,n+1).$
\newline
Also consider (6.21) as follows:
\begin{eqnarray}
\nonumber
R_{(\mathbf{A})(\mathbf{B})(\mathbf{C})(\mathbf{D})}=\\
\nonumber
=K[\eta_{(\mathbf{A})(\mathbf{D})}\eta_{(\mathbf{B})(\mathbf{C})}-\eta_{(\mathbf{A})(\mathbf{C})}\eta_{(\mathbf{B})(\mathbf{D})}]\\
\end{eqnarray}
in which $((A),(B),(C),(D))=(1,2,3,.....,n,n+1).$
As all points on the n+1-dimensional pseudo-sphere are geometrically equivalent,
we can see (6.41) as (6.40) in the origin of the coordinates frame $ X^{\Lambda}=0 $
or as the vielbein components at any point on the n+1-dimensional pseudo-sphere.
We conclude that a free classical massive particle in an n-dimensional pseudo-sphere
has classical motion and classical angular momentum constrained to the n-dimensional
pseudo-sphere, but has quantum momentum and quantum angular momentum in an
n+1-dimensional pseudo-sphere. An isotropic, homogeneous n-dimensional pseudo-sphere
has an infinite number of Ricci directions. Thus, from the classical point of
view, there is uncertainty in the particle position as well as in relation to the
classical angular momentum. It is always possible to go back to the classical original
metric, removing the classical particle uncertainty. From the quantum point of
view, there is an extra dimension, increasing the particle uncertainty in position,
quantum momentum and quantum angular momentum.
\newline
From the geometric point of view, some operations with the quantum angular momentum,
such as sums and products, suggest the same operations with curvature. We define some
procedures in differential geometry by operations with quantum angular momentum. For
example, consider the algebra of the group SO(p,n-p) given by (6.24), in which
\begin{eqnarray}
\nonumber
[L_{(\mathbf{A})(\mathbf{B})},L_{(\mathbf{C})(\mathbf{D})}]=-i(\eta_{(\mathbf{A})(\mathbf{C})}L_{(\mathbf{B})(\mathbf{D})}+\eta_{(\mathbf{A})(\mathbf{D})}L_{(\mathbf{C})(\mathbf{B})}\\
\nonumber +\eta_{(\mathbf{B})(\mathbf{C})}L_{(\mathbf{D})(\mathbf{A})}+\eta_{(\mathbf{B})(\mathbf{D})}L_{(\mathbf{A})(\mathbf{C})}).\\
\end{eqnarray}
We can place (6.19) in (6.24), obtaining a representation of the algebra in terms of
curvature operators. Placing (6.19) in (6.25), in terms of the curvature
operators, we have the form-invariance of the metric tensor $g'$ in relation to the algebra
so(p,n-p). Any other possible operation among quantum angular momenta defined herein
can be placed in terms of curvature. This offers some curious procedures in
differential geometry by simple operations with quantum angular momentum, which may
not be possible or are very difficult using geometric methods. The association between the
quantum angular momentum and differential geometry can be useful in both geometry
and physics.
\begin{center}
\section{Physical Principles Based on Geometric Properties}
 $                       $
\end{center}
 \setcounter{equation}{0}
 $         $
\newline
In this section, a new postulate is announced and some
physical principles are developed. However, we must first to make some considerations regarding
the electromagnetic field on an n-dimensional pseudo-sphere.
\newline
Consider the anti-symmetric electromagnetic of second rank tensor $F_{\alpha\sigma}$.
In relation to a Killing vector $\xi,$  the Lie derivative of
$g'^{\alpha\beta} F_{\alpha\sigma}=0$ is given by
\begin{eqnarray}
  L{_{\xi}({g'^{\alpha\beta}F_{\alpha\beta}}})=g'^{\alpha\beta}L{_{\xi}({F_{\alpha\beta}}})=0.
\end{eqnarray}
or, more explicitly,
\begin{eqnarray}
 L{_{\xi}({g'^{\alpha\beta}F_{\alpha\beta}}})=g'^{\mu\nu}[\frac{\partial{{\xi}^{\rho}}}{\partial{x^{\mu}}}F_{\rho\nu}+ \frac{\partial{{\xi}^{\rho}}}{\partial{x^{\mu}}}F_{\mu\rho}+
\xi^{\lambda}\frac{\partial{F_{\mu\nu}}}{\partial{{\xi}^{\lambda}}}]=0.
\end{eqnarray}
Using the symmetry of $g'^{\mu\nu}$, we have
\begin{eqnarray}
 L{_{\xi}({g'^{\alpha\beta}F_{\alpha\beta}}})=\frac{1}{2}
 .g'^{\mu\nu}[\xi^{\lambda}(\frac{\partial{F_{\mu\nu}}}{\partial{{\xi}^{\lambda}}}+\frac{\partial{F_{\nu\mu}}}{\partial{{\xi}^{\lambda}}})]=0,
\end{eqnarray}
or
\begin{eqnarray}
 L{_{\xi}({g'^{\alpha\beta}F_{\alpha\beta}}})=g'^{\mu\nu}[\xi^{\lambda}\frac{\partial{(0)}}{\partial{{\xi}^{\lambda}}}]=0.
\end{eqnarray}
 From (7.1) and (7.4), we have
\begin{eqnarray}
 L{_{\xi}({F_{\alpha\beta}}})=0.
\end{eqnarray}
Then, in relation to a Killing vector $\xi,$ $ F_{\alpha\beta}$ is maximally form-invariant.
\newline
A Killing vector $\xi$ obeys
\begin{eqnarray}
 \xi_{\mu;\nu}+\xi_{\nu;\mu}=0.
\end{eqnarray}
 After some considerations, placing (7.6) in (7.5)  we have \cite{13}
\begin{eqnarray}
(n-2)F_{\alpha\beta}=0,
\end{eqnarray}
which, for $n>2,$ implies
\begin{eqnarray}
F_{\alpha\beta}=0.
\end{eqnarray}
Thus, electric and magnetic fields vanish on an n-dimensional pseudo-sphere.
From the usual expressions of electric and magnetic fields, such as functions of
scalar and vector potentials, we have
\begin{eqnarray}
A_{\mu}=0.
\end{eqnarray}
\newline
Consider a density of electrical charges on an n-dimensional pseudo-sphere.
As all points of the sphere are geometrically equivalent, a density of
electrical charges has the same value at all points. Thus, it is a constant.
\newline
From Maxwell's equations, we have
\begin{eqnarray}
F^{\rho\sigma}{_{;\rho}}=-J^{\sigma}.
\end{eqnarray}
Placing (7.8) in (7.10), we obtain
\begin{eqnarray}
 J^{\sigma}=0.
\end{eqnarray}
In (5.22), we conclude that the sum of all charges is zero and the sum of all currents is also zero.
\newline
Consider, on an n-dimensional pseudo-sphere, a maximally form-invariant second rank tensor $B_{\alpha\sigma}$ as (7.5), with non-defined symmetry,
\begin{eqnarray}
 L{_{\xi}({B_{\alpha\beta}}})=0.
\end{eqnarray}
\newline
After some considerations, we have \cite{13}
\begin{eqnarray}
(n-2)(B_{\alpha\beta}-B_{\beta\alpha})=0,
\end{eqnarray}
which, for $n>2,$ implies
\begin{eqnarray}
B_{\alpha\beta}=B_{\beta\alpha}=const.g_{\beta\alpha}.
\end{eqnarray}
The only maximally form-invariant tensor of second rank different from zero is the metric tensor times a constant \cite{13}.
\newline
To obtain the quantum angular momentum from geometric considerations, we have
considered only constant vectors in a flat n+1-dimensional manifold.
We now reconsider the qualitative analyses of (6.9) made in
Section $6$. In the region in which $<U,N>$ does not vanish, $\bar{U}$ is a
conformal Killing vector and, in the region in which it vanishes, $\bar{U}$ is a
Killing vector. In other words, we have Killing and conformal Killing vectors
residing on an n-dimensional pseudo-sphere. For our purposes, we need only Killing
objects as the quantum angular momentum. Next, we present the following postulate:
\newline
\textbf{In (5.22) Nature always chooses Killing objects.}
\newline
Based on this postulate, we build four classical principles, one of which
is identified as a classical version of Heisenberg's
uncertainty principle and another is identified as a classical version of Bohr's
non-radiation postulate. The third principle is not new and is associated
to the electrical neutrality of a stable system. The fourth can be
interpreted as an equivalence between two descriptions of the
motion of the particle: The first as motion due to the presence
of forces and the second as a consequence of geometry, as in
Einstein's gravitation. For such, we assume only constant vectors in
an n+1-dimensional flat manifold, in which (5.22) is embedded.
\newline
Equations (2.28) and (2.34) tell us that the curvature is
null at the points in which $ A_{(\mathbf{A}){(\mathbf{C})(\mathbf{D})}}$ and
$B_{(\mathbf{A}){(\mathbf{B})(\mathbf{C})(\mathbf{D})}}$ are null.
In this case, the classical angular momenta are unspecified.
We conclude that any free particle in a curved manifold is always
in motion, with angular momentum not null, regardless of whether or
not we consider a physical theory.
\newline
Equation (5.24) tells us that $ ds^2$ is conformal to a flat
manifold and to (5.22). An observer in (5.22) will see the space as
being homogeneous and isotropic in the small region in which the
transformation (2.4) is well-behaved. Under this condition, Ricci
principal directions of space are indeterminate; so that
the position of the particle in this region is uncertain. In the conjugate
points of Jacobi's fields, the transformation (2.4) fails because
geodesics cross, mix or come into contact one other. Therefore, close to a
conjugate point, we will not have indetermination in the Ricci
principal directions and the uncertainty in the particle position disappears.
If (5.22) is valid at all points of the space, there will be an
indetermination of Ricci principal directions at each point and, consequently, a total
uncertainty regarding the position of the particle. This resembles a property of
Heisenberg's uncertainty principle and could be seen as a classical version.
\newline
The metric (5.22) is form-invariant for a displacement $ \xi $,
which is a Killing vector. From the postulate above, there are only
Killing vectors. In this metric, a physical scalar
field is constant or zero and anti-symmetric second rank
tensors, such as the electromagnetic tensor, are null. Under these conditions,
electromagnetic fields are trivial and there is no radiation.
In the neighborhood of the conjugate points,
normal coordinate transformations fail and there is no
indetermination of the Ricci principal directions. Moreover, the electromagnetic
fields are not trivial, being a radiative field. This is similar
to Bohr's postulate for radiation and could be seen as a classical
version.
\newline
In the region in which the transformation (2.4) is well-behaved, the metric
can be put in the form (5.22) and particles are in free motion without forces.
\newline
We can consider this a principle, creating an equivalence between two
descriptions of motion that are possible through normal transformations.
The first description, in local coordinates, is related to the conception of force.
The second is related to the conception of motion without force.
\newline
We believe that this principle moves in the direction of Einstein's dream,
as it points to the possibility of thinking in physics without forces,
as in Einstein's gravity.
\newline
We notice that the conjugate points of Jacobi's fields may be a consequence of
geometric singularities, as in the origin of Schwarzschild's geometry
\cite{14}, in which the curvature diverges. However, it may be due to the construction
of the coordinates, as in the case of a maximally symmetric space,
in which the curvature is finite at all points. In the latter case, we
have an indetermination of the Ricci principal directions, whereas there is no indetermination in the
former.
\newline
We recall that, in the region in which there are no conjugate points of
Jacobi's fields, it is possible to build a transformation (2.4) between
the ordinary metric and (5.24) as well as a conformal transformation between
(5.24) and (5.22). Because $g'$ is form-invariant in the region in which (5.22)
is defined, there are no fields or radiation. The quantum angular momentum,
which is a Killing object and resides in an n+1-dimensional pseudo-sphere,
appears as a geometric consequence of embedding (5.22) in a flat
n+1-dimensional manifold. Particles will be in a free motion, but, from
the classical point of view, confined on an n-dimensional pseudo-sphere,
and, from a quantum point of view, confined on an n+1-dimensional pseudo-sphere.
In this context, in which forces do not exist, particle confinement
is due to the manifold geometry. This resembles Einstein's geometric
vision as well as Heisenberg's uncertainty principle in quantum mechanics.
\section{Local Embedding of Riemannian Manifolds in Flat Manifolds}
 $        $
 \setcounter{equation}{0}
 $         $
In section 3, we presented considerations on the regions
in which coordinate transformations are well-defined.
\newline
Let us rewrite (3.10)
\begin{equation}
ds^2=\exp(2\sigma)\eta_{(\mathbf{A})(\mathbf{B})}dz^{(\mathbf{A})}dz^{(\mathbf{B})}.
\end{equation}
Defining the transformation of coordinates \cite{15},
\begin{equation}
y^{(\mathbf{A})}=\exp(\sigma)z^{(\mathbf{A})},
\end{equation}
with $ (A)=(1,2,3,....,n),$
\begin{equation}
y^{n+1}=\exp(\sigma)(\eta_{(\mathbf{A})(\mathbf{B})}z^{(\mathbf{A})}z^{(\mathbf{B})}-\frac{1}{4}),
\end{equation}
and
\begin{equation}
y^{n+2}=\exp(\sigma)(\eta_{(\mathbf{A})(\mathbf{B})}z^{(\mathbf{A})}z^{(\mathbf{B})}+\frac{1}{4}).
\end{equation}
It is easy to determine that
\begin{equation}
\eta_{\mathbf{A}\mathbf{B}}y^{\mathbf{A}}y^{\mathbf{B}}=0,
\end{equation}
in which
\begin{equation}
\eta_{\mathbf{A}\mathbf{B}}=(\eta_{(\mathbf{A})(\mathbf{B})},\eta_{\mathbf{(n+1),}\mathbf{(n+1)}},\eta_{\mathbf{(n+2),}\mathbf{(n+2)}}),
\end{equation}
with
\begin{equation}
\eta_{\mathbf{(n+1),}\mathbf{(n+1)}}=1,
\end{equation}
and
\begin{equation}
\eta_{\mathbf{(n+2),}\mathbf{(n+2)}=-1}.
\end{equation}
By a simple calculation, we find that the line element is
given by
\begin{equation}
ds^2=\exp(2\sigma)\eta_{(\mathbf{A})(\mathbf{B})}dz^{(\mathbf{A})}dz^{(\mathbf{B})}=
\eta_{\mathbf{A}\mathbf{B}}dy^{\mathbf{A}}dy^{\mathbf{B}}.
\end{equation}
Equation (8.5) is a hyper-cone in the (n+2)-dimensional flat
manifold. The metric (8.1) was embedded in the hyper-cone (8.5) of
the (n+2)-dimensional flat manifold.

\end{document}